\documentclass[pdflatex,sn-apa]{sn-jnl}

\usepackage{amsmath}
\usepackage{graphicx}
\usepackage{booktabs}
\usepackage{tabularx}
\usepackage{tikz}
\usetikzlibrary{positioning,arrows.meta}
\usepackage{hyperref}
\hypersetup{hidelinks,
  pdftitle={Validation Evidence in LLM Repair Agents: How Much of What Passes Actually Tests the Bug?},
  pdfauthor={Xiaonan Xu and Wenjing Wu}}

\AtBeginDocument{\let\appendices\oldappendices\let\endappendices\oldendappendices}

\DeclareRobustCommand{\$}{\ensuremath{\mathdollar}}

\newcommand{\SQDesignTasks}{110}
\newcommand{\SQDesignTasksPerSource}{55}
\newcommand{\SQDesignPlannedRollouts}{660}
\newcommand{\SQDesignReps}{2}
\newcommand{\SQPowerMcidCalibratedJoint}{91.2}
\newcommand{\SQPowerMcidEic}{96.2}
\newcommand{\SQPowerMcidBd}{93.2}
\newcommand{\SQPowerFiveppJoint}{28.8}
\newcommand{\SQDesignMcidAbsPp}{10}
\newcommand{\SQSFiveObservedRollouts}{643}
\newcommand{\SQSFiveMissingRollouts}{17}
\newcommand{\SQSFiveMissingBaseline}{6}
\newcommand{\SQSFiveMissingStatic}{5}
\newcommand{\SQSFiveMissingBcf}{6}
\newcommand{\SQSFiveRosterRepos}{64}
\newcommand{\SQSFivePreflightReplacements}{16}
\newcommand{\SQSFiveEicDiffPp}{$-$7.8}
\newcommand{\SQSFiveEicP}{0.0029}
\newcommand{\SQSFiveEicCiPp}{[$-$12.9, $-$2.7]}
\newcommand{\SQSFivePairedTasks}{109}
\newcommand{\SQSFiveBdDiffPp}{+7.4}
\newcommand{\SQSFiveBdP}{0.011}
\newcommand{\SQSFiveBdCiPp}{[+1.7, +13.0]}
\newcommand{\SQSFiveEicDiffAbsPp}{7.8}
\newcommand{\SQSFiveBdDiffAbsPp}{7.4}
\newcommand{\SQSFiveEicWildP}{0.00020}
\newcommand{\SQSFiveBdWildP}{0.0018}
\newcommand{\SQSFiveWildDraws}{99,999}
\newcommand{\SQSFiveSignEicBetter}{19}
\newcommand{\SQSFiveSignEicWorse}{5}
\newcommand{\SQSFiveSignEicP}{0.0066}
\newcommand{\SQSFiveSignBdBetter}{21}
\newcommand{\SQSFiveSignBdWorse}{7}
\newcommand{\SQSFiveSignBdP}{0.013}
\newcommand{\SQSFiveLoroRepos}{63}
\newcommand{\SQSFiveLoroEicFav}{63}
\newcommand{\SQSFiveLoroBdFav}{63}
\newcommand{\SQSFiveEicAdverseBoundPp}{$-$5.0}
\newcommand{\SQSFiveBdAdverseBoundPp}{+4.5}
\newcommand{\SQSFiveEicCompletePairsDiffPp}{$-$8.0}
\newcommand{\SQSFiveCompletePairsTasks}{100}
\newcommand{\SQSFiveNomvDiffPp}{+0.4}
\newcommand{\SQSFiveResolvedDiffPp}{+0.5}
\newcommand{\SQSFiveBcfbaseEicPp}{$-$10.5}
\newcommand{\SQSFiveBcfbaseBdPp}{+10.9}
\newcommand{\SQSFiveStaticEicAbsPp}{3.2}
\newcommand{\SQSFiveStaticBdAbsPp}{4.1}
\newcommand{\SQSFiveBaselineEicPct}{23.8}
\newcommand{\SQSFiveBaselineBdPct}{72.9}
\newcommand{\SQSFiveTriggerStaticPct}{97.2}
\newcommand{\SQSFiveTriggerBcfPct}{96.7}
\newcommand{\SQSFiveEventsTotal}{3,730}
\newcommand{\SQSFiveRoleRegressionOnly}{1,141}
\newcommand{\SQSFiveRoleGoldAlignedBugDiscriminating}{1,007}
\newcommand{\SQSFiveRoleCandidateSpecific}{370}
\newcommand{\SQSFiveRoleMisleading}{30}
\newcommand{\SQSFivePositiveComparable}{2,548}
\newcommand{\SQSFiveNondiscPositivePct}{46.0}
\newcommand{\SQSFiveBfailSpass}{1,377}
\newcommand{\SQSFiveCandSharePct}{26.9}
\newcommand{\SQSFiveCoexistPct}{48.1}
\newcommand{\SQSTwoAuditEvents}{111}
\newcommand{\SQSTwoAuditDisagreements}{8}
\newcommand{\SQSTwoAuditAccuracyPct}{92.8}
\newcommand{\SQSSixMissingRollouts}{12}
\newcommand{\SQSSixMissingArmSplit}{4/4/4}
\newcommand{\SQSSixPygmtCells}{6}
\newcommand{\SQSFiveSameTwoZeroBaselineEicPct}{18.9}
\newcommand{\SQSSixBaselineEicPct}{8.3}
\newcommand{\SQSSixbMissingRollouts}{7}
\newcommand{\SQSSixbMissingArmSplit}{3/2/2}
\newcommand{\SQSSixbTriggerPct}{97.4}
\newcommand{\SQSSixbBaselineEicPct}{40.5}
\newcommand{\SQSSixbPlaceboEicAbsPp}{7.5}
\newcommand{\SQSFiveBreplayMedianS}{11.0}
\newcommand{\SQSFiveBreplayWallPct}{3.3}
\newcommand{\SQSFiveTokensDiff}{99}
\newcommand{\SQRqTwoN}{643}
\newcommand{\SQRqTwoRepositories}{64}
\newcommand{\SQRqTwoFolds}{ten}
\newcommand{\SQRqTwoBrierDesc}{0.231}
\newcommand{\SQRqTwoBrierFull}{0.203}
\newcommand{\SQRqTwoBrierReduction}{0.028}
\newcommand{\SQRqTwoBrierReductionCi}{[0.003, 0.053]}
\newcommand{\SQRqTwoAurocDesc}{0.521}
\newcommand{\SQRqTwoAurocFull}{0.694}
\newcommand{\SQRqTwoCalibIntercept}{0.22}
\newcommand{\SQRqTwoCalibSlope}{0.63}
\newcommand{\SQCMaxOutputTokens}{6{,}000}
\newcommand{\SQCGuardResponses}{200}
\newcommand{\SQCGuardToolCalls}{200}
\newcommand{\SQCReplTasks}{20}
\newcommand{\SQCReplPlannedCells}{120}
\newcommand{\eic}{EIC}

\begin{document}

\title[Validation evidence in LLM repair agents]{Validation Evidence in LLM Repair Agents: How Much of What Passes Actually Tests the Bug?}

\author*[1]{\fnm{Xiaonan} \sur{Xu}}\email{xiaonanxu5@gmail.com}
\author[2]{\fnm{Wenjing} \sur{Wu}}
\affil*[1]{\orgdiv{College of Computing}, \orgname{Georgia Institute of Technology}, \orgaddress{\city{Atlanta}, \state{GA}, \postcode{30332}, \country{USA}}}
\affil[2]{\orgdiv{Department of Computer Science}, \orgname{University of Colorado Boulder}, \orgaddress{\city{Boulder}, \state{CO}, \postcode{80309}, \country{USA}}}

\abstract{When a repair agent runs a test and sees it pass, the result is
treated as evidence about the reported defect. We measure how often that
treatment is warranted. BSG-VA (buggy-state/candidate-state/gold-fix
validation analysis) captures each validation command at its exact
working-tree state, extracts a test-only patch, and replays the command on
the original buggy code (B), the candidate state (S), and the developer gold
fix (G). The captured outcome and the replay results assign every event an
evidence role, from
gold-aligned bug-discriminating through regression-only to misleading. Across
\SQSFiveEventsTotal{} events in \SQSFiveObservedRollouts{} rollouts on
\SQDesignTasks{} tasks, \SQSFiveNondiscPositivePct\% of positive comparable
events carry no bug-discriminating information;
\SQSFiveBaselineEicPct\% of baseline rollouts, with no feedback injected,
close with a patch whose entire positive evidence base is of this kind. A three-arm experiment tests
whether returning the B-replay outcome to the agent changes this pattern.
Bug-contrast feedback reduces evidence-inadequate closure by
\SQSFiveEicDiffAbsPp{} percentage points relative to an attention-matched
reminder ($p = \SQSFiveEicP$) and raises bug-discriminating evidence by
\SQSFiveBdDiffAbsPp{} points ($p = \SQSFiveBdP$), with no detectable cost to
repair success. Both estimates fall below the prespecified
\SQDesignMcidAbsPp-percentage-point smallest effect size of interest, so
practical magnitude remains uncertain. Roughly a third of the improvement traces
to the reminder alone; across two exploratory replications, varying the scaffold
and the model, the B-replay content adds a detectable increment only with
gpt-5.6-sol under the unconstrained tool-use loop.
BSG-VA applies post hoc to any replayable repair trajectory that
preserves the required code states and execution environment.}


\maketitle

\section{Introduction}\label{sec:intro}

Repair agents on SWE-bench-style benchmarks now resolve a substantial and
growing share of real-world GitHub issues. A typical successful trajectory
involves dozens of tool calls: reading files, editing code, running tests,
observing outputs, and iterating. Among these actions, running a test and
seeing it pass occupies a privileged position: it is the closest thing the
agent has to empirical confirmation that its changes work. Benchmark
evaluation reinforces the reading, since the final verdict is whether the
submitted patch passes a held-out test suite.

But a passing test can mean different things. Consider an agent assigned a
defect in a date-formatting routine. The agent modifies the routine, then
writes a test that imports the module and asserts that the output is a string.
The test passes. It would also have passed on the original buggy code, because
the bug was not about return types but about locale handling. The agent has
confirmed that its patch does not break the import or change the output type.
It has learned nothing about whether the locale bug is fixed. If this is the
only positive evidence the agent collects before submitting, the submission
rests on validation that is real but irrelevant to the assigned defect.

How common is this pattern? No existing method answers the question, because
mid-trajectory validation events have not been systematically evaluated for
evidential content. Prior work on agent-generated tests has treated them as
standalone artefacts: the SWT-Bench family evaluates dedicated pipelines that
take an issue description and produce a test designed to fail on buggy code
and pass on the fix \citep{SWT,OTTER,ASSERTFLIP}. These are dedicated
systems, separate from the repair process. Trajectory analyses of frontier
models \citep{RETHINK} have characterised the frequency and syntactic form of
in-trajectory test events, showing that agents write tests often but not
evaluating what those tests prove about the assigned bug. The gap between
``the agent ran a test'' and ``the test targeted the defect'' remains
unmeasured.

We introduce BSG-VA (buggy-state/candidate-state/gold-fix validation analysis)
to close this gap. The method intercepts every validation command the agent
executes, snapshots the working tree at execution time, and extracts a
test-only patch that separates the validation logic from any concurrent
production edits. It then replays the same command on three code states: the
original buggy version (B), the captured candidate (S), and the developer gold
fix (G). The captured outcome and the replay results place each event in one
of seven evidence roles. An event where the test fails on B, passes on S, and
passes on G is gold-aligned bug-discriminating: it targets a property specific
to the defect, in a way that would also accept the developer's own fix. An
event that passes on B as well as S is at best regression-only: it confirms
the patch has not introduced a new failure, but says nothing about the bug.
Between these poles sit candidate-specific events (B-fail, S-pass, G-fail),
misleading events, and diagnostic negatives; two further categories set aside
events whose outcomes are unstable or not evaluable. The taxonomy is
exhaustive, mutually exclusive, and defined entirely by recorded, observable
outcomes.

We apply BSG-VA at scale in a controlled experiment using gpt-5.6-sol
(hereafter \emph{sol}) on \SQDesignTasks{} tasks drawn equally from SWE-bench Verified and SWE-rebench.
The measurement covers \SQSFiveEventsTotal{} retained post-edit validation
events across \SQSFiveObservedRollouts{} rollouts (events before the agent's
first production edit are excluded, as discussed in
Section~\ref{sec:method}). Of \SQSFivePositiveComparable{} positive
comparable events, \SQSFiveNondiscPositivePct\% are regression-only or
misleading. At the rollout level, \SQSFiveBaselineEicPct\% of baseline runs
exhibit what we call evidence-inadequate closure: the agent submitted a patch,
collected positive validation results, and none of those results discriminated
the reported bug from a regression-free alternative.

The measurement enables a direct intervention. Because BSG-VA can evaluate
an event's B-replay outcome in real time (median \SQSFiveBreplayMedianS{}
seconds per rollout), we test whether feeding that outcome back to the agent
changes the closure pattern. In the three-arm experiment, bug-contrast feedback
(BCF) takes the agent's passing validation event, replays it on B, and reports
the result in a system message. If the test also passes on B, the agent learns
its evidence is non-discriminating. The comparison arms are a
structure-matched generic reminder that prompts the agent to reconsider its
evidence without providing B-replay information, and a baseline arm into
which no messages are injected.
BCF reduces evidence-inadequate closure by \SQSFiveEicDiffAbsPp{} percentage
points relative to the reminder ($p = \SQSFiveEicP$, 95\% CI \SQSFiveEicCiPp)
and raises bug-discriminating evidence by \SQSFiveBdDiffAbsPp{} points
($p = \SQSFiveBdP$). Repair success shows no detectable change. The active-control
decomposition shows the generic reminder alone accounts for roughly a third of
the total BCF-versus-baseline improvement, a share that comes from prompting
the agent to attend to evidence quality rather than from the B-replay content
itself.

Under a plan-execute-verify scaffold, baseline evidence-inadequate closure
falls to \SQSSixBaselineEicPct\%, a floor effect that leaves little room for
either component. On gpt-5.6-terra (hereafter \emph{terra}) in the
unconstrained loop, the generic reminder
produces a \SQSSixbPlaceboEicAbsPp-point reduction on its own, comparable to
the full BCF effect on sol, while the B-replay content adds nothing
detectable. The prompting effect recurs on the second model; the B-replay
content adds measurable value only with gpt-5.6-sol in the unconstrained
loop.

The study is organised around four objectives. RQ1 (descriptive): what share
of positive validation evidence produced mid-trajectory discriminates the
assigned bug? RQ2 (associational): does evidence quality carry predictive
signal for official resolution (the benchmark's own pass/fail verdict)
beyond task-level covariates? The confirmatory
objective tests whether BCF reduces evidence-inadequate closure relative to a
structure-matched generic reminder, with bug-discriminating evidence as a
gated secondary outcome. Two exploratory replications \citep{SHULL2008} then
ask how these patterns change across a different scaffold and a second model.

This paper contributes an event-level measurement method and evidence-role
taxonomy that assign each validation event an evidence role with directly
checkable content; a quantified prevalence estimate showing that nearly
half of the positive comparable validation evidence in a
\SQSFiveEventsTotal-event study does not discriminate the assigned bug; and a
fully crossed three-arm experiment that separates the prompting effect of a
generic reminder, which recurs on a second model, from the additional
increment of B-replay content, which does not.

\section{Background and related work}\label{sec:related}

\subsection{Repository-level program repair with LLM agents}

Automated program repair (APR) has evolved from template-based and search-based
patch generation \citep{GOUES2012,WEIMER2009} toward neural and
large-language-model approaches \citep{XIA2023,JIANG2023}. The
current generation combines a large language model with an agentic tool-use
loop: the model reads an issue description, navigates the repository, edits
files, and runs commands, iterating until it produces a candidate patch or
exhausts a budget. SWE-agent \citep{SWEAGENT} established the architecture;
Agentless \citep{AGENTLESS} showed that a fixed pipeline of
localisation, repair, and patch validation without persistent agent state can
be competitive; CodeR \citep{CODER},
Moatless \citep{MOATLESS}, and AutoCodeRover \citep{AUTOCODEROVER} introduced
variations on retrieval, planning, and tool design. SWE-bench \citep{SWEBENCH}
and its curated subset SWE-bench Verified \citep{SWEBENCH_VERIFIED} provide the
standard evaluation protocol: the agent receives an issue, works within a
repository checkout, and submits a patch that is evaluated against a held-out
developer test suite.

Across all these architectures, agents routinely execute validation commands
during the trajectory. The evaluation protocol, however, examines only the
final submitted patch. The mid-trajectory validation activity is logged but otherwise
ignored by the benchmark harness.

\subsection{Deliberately generated tests for patch validation}

A separate body of work generates tests deliberately, as a post-hoc step to
validate candidate patches. SWT-Bench \citep{SWT} defines the fail-to-pass
criterion: a generated test must fail on the original buggy code and pass on
the developer fix. Filtering agent patches through such tests raises precision
substantially \citep{SWT}. The protocol has become a competitive benchmark in
its own right, with systems including Otter, AssertFlip, Issue2Test, Echo, and
EvoOtter optimising fail-to-pass success rates
\citep{OTTER,ASSERTFLIP,ISSUE2TEST,ECHO,EVOOTTER}, and with repair agents
tasked to co-generate fixes and reproduction tests inside the repair loop
\citep{DYNAMIC_BRT}. LIBRO \citep{LIBRO} generates issue-reproducing tests
from bug reports without access to the developer fix at generation time. A
valid fail-to-pass test is still not a complete oracle: it can cover a single
manifestation of the issue and steer repair toward a partial patch
\citep{SWEDOCTOR}. Assessment of generated suites has its own tooling, from
mutation-based measures of discriminative power \citep{SWE_MUTATION} to
lifecycle evidence infrastructure that tracks execution, coverage, and
flakiness \citep{TESTMAP}.

The distinction between this body of work and the present study is one of
context and agency. Deliberate test generation is an external pipeline step,
executed after the agent's trajectory is complete, with the explicit goal of
producing a discriminating test. BSG-VA examines the agent's own spontaneous
validation behaviour during the trajectory, where the agent decides what to
test, when, and how, without any external instruction to produce a
fail-to-pass test. The two perspectives are complementary: the former asks
whether discriminating tests can be generated on demand, while the latter asks
what the tests that agents already generate actually prove.

\subsection{Trajectory analysis and test-value assessment}

\citet{RETHINK} analyse six models on SWE-bench Verified and report that
test-writing frequency is similar between successful and failed trajectories;
that most validation commands are print-statement explorations rather than
assertion-bearing tests; and that prompting models to write more or fewer
tests produces no detectable change in repair outcomes. The finding that
test-writing is abundant but not predictive of success is consistent with the
hypothesis that much of the validation activity lacks discriminating power,
but the study does not measure discriminating power directly. \citet{TRNR}
analyse the cost-effectiveness of code execution in LLM-based repair at scale
and report that failed commercial-agent runs often pass their own
self-validation while failing official evaluation. That mismatch is the
starting point here: where their analysis works at the level of run frequency,
cost, and coarse agreement between the agent's own test results and the
official evaluation, BSG-VA reconstructs the exact candidate state of each
self-selected command and replays
it to establish which passing checks could not have detected the defect.

Broader diagnostics work instruments trajectories at a structural level:
lucky-pass detection and production trajectory review
\citep{AGENTLENS_LUCKY,AGENTLENS_PROD}, trajectory-structure and
verification-skip diagnostics \citep{TRACEPROBE,TAR_TRAJ}, traceability
analyses of reproduction and regression testing across repair agents
\citep{TRACEABILITY}, and interactive trajectory visualisation
\citep{TRACEVIEW}. Cross-framework comparison warns that the same aggregate
trajectory signal can carry different semantics in different frameworks
\citep{SAME_SIGNAL}. BSG-VA is narrower: it attaches an executable,
issue-specific meaning to individual validation events rather than
characterising trajectory structure.

Overfitting to the validation signal has a long history in APR: generated
patches can overfit weak test suites \citep{SMITH2015}, and test generation
was proposed early to expose overfitted patches \citep{DIFFTGEN}.
\citet{CODING_BEFORE_TEST} approach the agent-era version of the problem,
showing that tests generated after exposure to an incorrect implementation
detect fewer faults than independently generated tests, a mechanism by which
non-discriminating suites arise. Controlled evidence that incorrect code
context reduces generated-test fault detection supports the same account
\citep{INCORRECT_CODE}. \citet{OVERMOCK} document a related fragility in
which agent commits mock more heavily than non-agent commits, risking tests
that compile and run but exercise little of the relevant behaviour.
\citet{B2T} show, in a controlled code-as-spec study, that agents can satisfy
every visible check without making the requested artefact load-bearing. These
findings identify specific mechanisms by which validation evidence can be
non-discriminating, but none provides a systematic taxonomy of evidence roles
or a prevalence estimate across a controlled population.

BSG-VA extends this line by operationalising the measurement. Rather than
classifying tests by static syntactic signals, as in large-scale oracle-signal
audits of agent-authored test code \citep{SMOKE}, or by comparing agent tests
with developer tests as monolithic suites, BSG-VA captures each event at its
exact state, isolates the test-only component, and replays it on the buggy,
candidate, and gold states. The captured outcome and the replay pattern assign
each individual event an evidence role with directly checkable content,
yielding an event-level taxonomy and prevalence estimate.

\subsection{Feedback and steering in agentic repair}

Existing work on steering agent behaviour during repair has focused on tool
design, retrieval augmentation, and planning structure
\citep{SWEAGENT,AUTOCODEROVER,MOATLESS}. At a conceptual level, automated
verification signals have been framed as imperfect proxies constrained by
scalability, faithfulness, and robustness \citep{VERIFY_HORIZON}, and
process-discipline benchmarks score whether agents verify and recover at all
\citep{RIGORBENCH}. Closest in spirit to an evidence-quality intervention,
EviACT gates repair actions behind evidence-driven retrieval, compilation, and
target-test guardrails \citep{EVIACT}. Few studies have investigated online
feedback about the quality of the agent's own validation evidence as a
steering signal. The approach tested here, in which the agent receives the
outcome of replaying its own test on the original buggy state, is related in
spirit to self-debugging techniques \citep{CHEN2023SELFDEBUG} that feed
execution output back into the model, but differs in that the information
provided is a counterfactual comparison (the test on a state the agent never
visited) rather than an observation of the current execution.

\section{The BSG-VA measurement method}\label{sec:method}

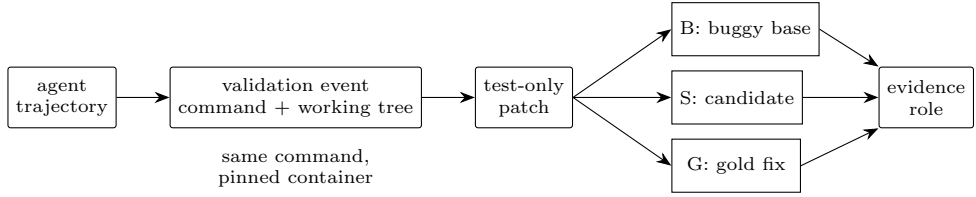
\begin{figure}[t]
\centering
\begin{tikzpicture}[
  node distance=4mm and 7mm,
  box/.style={draw, rounded corners=1pt, align=center, font=\footnotesize, inner sep=3pt, minimum height=8mm},
  st/.style={draw, align=center, font=\footnotesize, inner sep=3pt, minimum width=17mm, minimum height=7mm},
  arr/.style={-{Stealth[length=2mm]}, thin}
]
\node[box] (traj) {agent\\trajectory};
\node[box, right=of traj] (ev) {validation event\\command + working tree};
\node[box, right=of ev] (tp) {test-only\\patch};
\node[st, right=13mm of tp, yshift=9mm] (B) {B: buggy base};
\node[st, right=13mm of tp] (S) {S: candidate};
\node[st, right=13mm of tp, yshift=-9mm] (G) {G: gold fix};
\node[box, right=10mm of S] (role) {evidence\\role};
\draw[arr] (traj) -- (ev);
\draw[arr] (ev) -- (tp);
\draw[arr] (tp.east) -- (B.west);
\draw[arr] (tp.east) -- (S.west);
\draw[arr] (tp.east) -- (G.west);
\draw[arr] (B.east) -- (role.north west);
\draw[arr] (S.east) -- (role.west);
\draw[arr] (G.east) -- (role.south west);
\node[font=\footnotesize, align=center, below=1.5mm of ev] {same command,\\pinned container};
\end{tikzpicture}
\caption{BSG-VA: each self-selected validation event is captured at its exact
working-tree state, reduced to a test-only patch, and replayed deterministically
on the buggy base (B), the captured candidate (S), and the developer gold fix
(G); the captured outcome and the replay pattern assign an evidence role}\label{fig:pipeline}
\end{figure}

\subsection{Event capture}

BSG-VA instruments the agent's tool loop so that every shell command the agent
issues after its first production edit is intercepted and classified
(Figure~\ref{fig:pipeline}). Classification is online and identical across all
experimental arms. Each command is labelled as one of four types: a
project test run (an invocation of the repository's own test runner on
its own test files), a scripted explicit oracle (an agent-written script whose
exit code encodes a pass/fail verdict), an observational probe (output to be
inspected by the model without a programmatic oracle), or a non-validation
action. Non-validation actions are logged but excluded from the measurement.
The classifier was audited against a sample of \SQSTwoAuditEvents{} events
labelled manually by an author; it agreed on
\SQSTwoAuditAccuracyPct\% of them, the \SQSTwoAuditDisagreements{} disagreements
were adjudicated by the second author, and all resulting corrections were
adopted into the production classifier.

For each validation event the harness records the command string, the standard
output and error streams, the exit status, and the working-tree state at
execution time. The working-tree state is decomposed into two components: a
production patch containing the agent's changes to the repository's source and
configuration files, and a test-only patch containing everything the agent
added or modified to support the validation itself (new test files, test
helpers, fixture data). The decomposition matches changed paths against the repository's existing test
directory conventions and the set of files the agent created during the
trajectory.

\subsection{Test-only patch extraction}

Reducing the captured state to a test-only patch is the step that makes
cross-state replay meaningful. If the full working tree were replayed on B,
any test depending on the agent's production changes would fail for trivial
reasons (missing functions, changed imports), and the replay outcome would
reflect dependency breakage rather than evidential content. The test-only
patch retains only the validation artefacts: the test files, scripts, and
fixtures the agent created or modified, applied on top of whatever code state
the replay targets. An event whose test-only patch is empty indicates that the
agent ran a pre-existing repository test without adding test code; these
events are retained because the choice of which existing test to run is itself
a validation decision.

The extraction handles several boundary cases. When the agent's test imports a
function that exists only in the agent's production patch, the replay on B
will fail with an import error rather than a test-logic failure. BSG-VA flags
such events as depending on a new production symbol. When the agent creates a
temporary file that its test reads, the file is included in the test-only
patch. When the agent modifies an existing test file, the extraction preserves
the modification against the repository's baseline version of that file. These
cases are defined in advance in the event codebook; the classification is
deterministic given the working-tree snapshot.

\subsection{Cross-state replay}\label{sec:replay}

Each retained event is replayed on three code states, each constructed in a
fresh container from the same base image used during the agent's original run.

State B is the original buggy code with the test-only patch applied. B
represents the question: would this validation event have detected the
reported defect? If the test passes on B, the event carries no information
that distinguishes buggy from candidate code.

State S is the captured candidate code. Replaying on S checks consistency: in
the common case the outcome matches what the agent observed during the
trajectory, and Section~\ref{sec:roles} defines how the remaining cases are
classified.

State G is the developer gold fix: the buggy code with the benchmark's gold
patch applied, plus the same test-only patch. G answers the question: is the event's discriminating power tied to the specific
candidate, or does it generalise to the correct fix? An event that fails on B
and passes on S but also fails on G discriminates the bug from this particular
candidate, but would reject the developer's own solution.

Every event is replayed once per state. Stability is assessed by
comparing the S-replay with the captured outcome: disagreement marks the event
flaky, and flaky events are excluded from all constructs. This rule favours
precision over recall: only stable observations support role assignment.

\subsection{Evidence-role taxonomy}\label{sec:roles}

Two kinds of outcome attach to each event. The captured outcome is the result
the agent observed during the trajectory. The replay outcomes are what the
offline replays produce on B, S, and G. Roles are assigned in a fixed order
that makes the seven categories mutually exclusive and exhaustive: the
captured outcome anchors the assignment, the B and G replays supply the bug
contrast, and the S-replay enters through the stability comparison of
Section~\ref{sec:replay}.

\texttt{DIAGNOSTIC\_NEGATIVE}. The captured outcome is not positive. The event
gave the agent no positive result on the candidate and carries no positive
evidence, whatever its replays show.

\texttt{FLAKY}. The captured outcome is positive but the S-replay disagrees
with it. The observation is not reproducible; flaky events are excluded from
all constructs.

\texttt{NOT\_COMPARABLE}. The captured outcome is positive and stable, but the
B-replay or the G-replay is not evaluable. Reasons include import errors from
tests that depend on symbols introduced by the agent's production edits,
missing temporary artefacts, container setup failures, and timeout.

Every remaining event has a stable positive captured outcome with evaluable B
and G replays and receives one of four roles.

\texttt{GOLD\_ALIGNED\_BUG\_DISCRIMINATING}. B fails, G passes. The event
detects a property present in the buggy code and absent in both the candidate
and the gold fix. This is the strongest form of validation evidence: it
targets the defect and generalises beyond the specific candidate.

\texttt{CANDIDATE\_SPECIFIC}. B fails, G fails. The event distinguishes the
bug from the candidate, but its discriminating power depends on implementation
details unique to the agent's patch. The gold fix does not satisfy this test.

\texttt{REGRESSION\_ONLY}. B passes, G passes. The event confirms only that
the candidate has not introduced a failure relative to the original code. It
says nothing about whether the bug is fixed.

\texttt{MISLEADING}. B passes, G fails. The check is one the buggy code
satisfies and the developer gold fix does not: the passing result contradicts
the developer's fix, so reading it as support for the candidate misleads.

An event is positive comparable when its captured outcome was positive and its
assigned role is one of the four roles from
\texttt{GOLD\_ALIGNED\_BUG\_DISCRIMINATING} to \texttt{MISLEADING}; under the
assignment order above, these are exactly the events that reach the four
outcome roles. These are the events the agent could have counted as supporting
evidence, evaluated for what they in fact establish.

Two rollout-level constructs aggregate the event roles. A rollout provides
bug-discriminating evidence when it contains at
least one gold-aligned bug-discriminating or candidate-specific event, the two
roles whose checks fail on the buggy state. Evidence-inadequate closure
(\eic) is true when a rollout submitted a patch, produced at least one
positive comparable event, and contains no bug-discriminating event: every
positive comparable result is regression-only or misleading.

\subsection{Use of AI assistants}\label{sec:ai-use}

An LLM assistant (Claude, Fable~5 model, Anthropic) was used in four
capacities: configuring the container-based experiment environment (Docker
images, harness wiring, replay infrastructure); implementing and running the
harness and orchestration code that launched experiment sessions and collected
data, under designs specified by the authors (the repair agents under study
are separate OpenAI models, not this assistant); writing analysis code; and
improving the readability and language of text written by the authors. All experimental
parameters, analysis plans, and decision rules were specified by the authors
before the corresponding model requests. The analysis code was verified by
the authors through code review, unit-level checks against manually computed
reference values, and reproduction of every reported statistic from the
released data. The authors reviewed all LLM-assisted text, revised it
against the style and accuracy standards of the study, and take full
responsibility for the final content.

\section{Study design}\label{sec:design}

\subsection{Task population}

We drew tasks equally from two sources: SWE-bench Verified, a curated subset
of SWE-bench with confirmed human solutions
\citep{SWEBENCH,SWEBENCH_VERIFIED}, and SWE-rebench, a dataset of GitHub
issues filed after the model's training cutoff, with lower contamination risk
\citep{SWEREBENCH}. The
two sources define two strata, one of curated verified tasks and one of recent
tasks with lower contamination risk, weighted equally in all
primary analyses. Each task specifies a repository, a commit, an issue
description, and a container image. Official resolution is judged by each
benchmark's evaluator on the submitted production patch, with agent-written
test artefacts stripped before evaluation.

\subsection{Three-arm design}

Every task is run under all three arms, with \SQDesignReps{} repetitions per
task and arm: a fully crossed, within-task design. Because each task serves as
its own control, no between-arm randomisation of tasks is required, and every
arm contrast is estimated within task by the estimator of
Section~\ref{sec:confirmatory-design}.

Baseline. The agent works with the default scaffold: an unconstrained tool-use
loop in which the model freely interleaves file reads, edits, and command
execution until it submits (hereafter \emph{the unconstrained loop}). No
messages are injected.

Static Reminder. After each eligible positive validation event, a system
message prompts the agent to reconsider whether its evidence targets the
reported defect. The message carries no B-replay information. Its structure,
length, and directive strength are matched to the BCF message, so this arm
serves as an attention control that isolates the effect of the B-replay content
from the effect of prompting the agent to attend to evidence quality.
Table~\ref{tab:messages} reproduces the text.

Bug-Contrast Feedback (BCF). After each eligible positive validation event,
the harness replays the event's test-only patch on B in real time and injects
the outcome as a system message. If the test passes on B, the message states
that the check does not distinguish the candidate from the original buggy
code. If the test fails on B, the message confirms that the check detects a
difference. Events where the B-replay is not evaluable receive a message
stating that the comparison was not possible. Table~\ref{tab:messages}
reproduces all message variants.

All arms share one model configuration: gpt-5.6-sol in standard reasoning mode
with high effort, \SQCMaxOutputTokens{} maximum output tokens per response,
and no repository network access. Each rollout is capped at
\SQCGuardResponses{} responses and \SQCGuardToolCalls{} tool calls.

Eligibility for feedback is defined identically in all arms: the event must be
a validation event with a positive captured outcome, and it must not be a
duplicate of an event already triggered in the same rollout. In the baseline
arm, eligible events are identified but no message is sent; this ensures that
trigger-rate statistics are comparable across arms.

\begin{table}[t]
\caption{Intervention messages. BCF delivers exactly one of
its three variants according to the deterministic B-replay outcome}\label{tab:messages}
\begin{tabular}{p{2.6cm}p{9.2cm}}
\toprule
Condition & Message \\
\midrule
Static Reminder & This check passed in the current candidate. A passing check may cover only part of the reported issue. Before concluding, consider whether it is sensitive to the reported behavior and whether regressions remain. \\
BCF, B passes & This check also passes on the original buggy version. It provides regression evidence but does not distinguish the two states. Before concluding, consider whether your validation is sensitive to the reported behavior and whether regressions remain. \\
BCF, B fails & This check fails on the original buggy version and passes in the current candidate. It distinguishes the two states, but one check does not establish complete correctness. Before concluding, consider remaining issue behavior and regressions. \\
BCF, not comparable & The original-version replay did not reach a comparable behavioral check, so the current pass is unverified evidence for the reported issue. Before concluding, consider a check that is directly sensitive to the reported behavior. \\
\botrule
\end{tabular}
\footnotetext{Messages are reproduced verbatim as delivered to the agent (hence the American spelling).}
\end{table}

\subsection{Confirmatory experiment}\label{sec:confirmatory-design}

The confirmatory experiment covers \SQDesignTasks{} tasks
(\SQDesignTasksPerSource{} per stratum) with \SQDesignReps{} repetitions of
each task under each arm, giving \SQDesignPlannedRollouts{} planned cells.
\SQSFivePreflightReplacements{} candidate tasks failed an automated preflight
environment check at startup and were replaced by the next candidate from the
same stratum before any model request, giving a final set of \SQDesignTasks{}
tasks across \SQSFiveRosterRepos{} repositories. Table~\ref{tab:design} summarises the design. The analysis plan, task
list, and decision rules were specified before any confirmatory model
request.

The primary outcome is \eic{} (BCF versus Static Reminder). The secondary
outcome, tested in a fixed sequence gated on the primary \citep{WK2001}, is
bug-discriminating evidence (BCF versus Static Reminder). The estimator is a
task-level paired difference: for each task with at least one observation in
both arms, we compute the within-task rate difference, average within each
stratum, and take the unweighted mean of the two stratum averages (the
\emph{equal-source} estimate, weighting the two strata equally). Inference
uses a two-stratum Welch t-test with Satterthwaite degrees of freedom
\citep{WELCH1947,SATTERTHWAITE1946}. The smallest effect size of interest
(SESOI) \citep{LAKENS2018} was set before execution to \SQDesignMcidAbsPp{}
percentage points: an effect of that size would change which of a repair
agent's closures an integrator would accept. The SESOI sized the experiment;
it is not a threshold for dismissing smaller observed effects. A
further secondary outcome, closure without positive machine-verifiable
evidence, monitors whether the feedback suppresses validation activity.

The analysis plan also includes a secondary associational
analysis (RQ2) examining the relationship between evidence quality and repair
success through logistic models with repository-grouped cross-validation,
computed from the released data with the plan's seed and procedure.

\begin{table}[t]
\caption{Confirmatory experiment population and design}\label{tab:design}
\begin{tabular}{lc}
\toprule
Tasks & \SQDesignTasks{} (\SQDesignTasksPerSource{} per stratum) \\
Repositories & \SQSFiveRosterRepos{} \\
Arms & Baseline, Static Reminder, BCF \\
Repetitions per task and arm & \SQDesignReps{} \\
Planned rollout cells & \SQDesignPlannedRollouts{} \\
Observed cells & \SQSFiveObservedRollouts{} \\
Cells lost to failures (by arm) & \SQSFiveMissingRollouts{} (\SQSFiveMissingBaseline/\SQSFiveMissingStatic/\SQSFiveMissingBcf{} by arm) \\
Model & gpt-5.6-sol, standard reasoning, high effort \\
Scaffold & unconstrained tool-use loop; no timeout, no automatic retry \\
Primary contrast & BCF $-$ Static Reminder \\
Primary outcome & \eic{} (favourable direction negative) \\
Key secondary outcome & bug-discriminating evidence (favourable direction positive) \\
\botrule
\end{tabular}
\end{table}

\subsection{Exploratory replications: varying the scaffold and the model}

Two exploratory replications probe generality on a shared
\SQCReplTasks-task subset selected by a deterministic outcome-blind rule from
the confirmatory task set. The scaffold replication was specified together with the
confirmatory design; the second-model replication was designed after the scaffold
results were available and specified before any request to the second model.

The scaffold replication replaces the unconstrained loop with a plan-execute-verify
scaffold that requires the agent to submit a plan before executing and gates
the final submission on a verification phase. The model remains gpt-5.6-sol.
Three arms, \SQDesignReps{} repetitions per task, \SQCReplPlannedCells{}
cells.

The second-model replication replaces gpt-5.6-sol with gpt-5.6-terra in the
unconstrained loop. Three arms,
\SQDesignReps{} repetitions, \SQCReplPlannedCells{} cells. Estimates from
both studies are reported with confidence intervals and descriptive p-values
and carry no confirmatory weight.

\subsection{Power and sample size}

We calibrated sample size to the prespecified SESOI of \SQDesignMcidAbsPp{}
percentage points rather than to the pilot point estimates, and the estimates
below are reported against it rather than filtered by it. No published
threshold was available to take the value from, because the construct is
introduced here; Section~\ref{sec:confirmatory-design} gives the design
judgement it rests on. Two pilots preceded the experiment; their data served only
to develop the measurement and intervention pipeline and to parameterise the
power model below, and pilot observations enter no confirmatory estimate.
Under a data-generating process using the pilot's stratum-level base rates
and within-task dependence, \SQDesignTasks{} tasks
with \SQDesignReps{} repetitions yield \SQPowerMcidEic\% power for \eic{} and
\SQPowerMcidBd\% for bug-discriminating evidence at the SESOI, and
\SQPowerMcidCalibratedJoint\% joint power. A supplementary
five-percentage-point scenario, added before execution, yields
\SQPowerFiveppJoint\% joint power, consistent with the expectation that an
effect below the SESOI would likely go undetected.

\section{Results}\label{sec:results}

\subsection{Evidence-role taxonomy at scale}\label{sec:prevalence}

\begin{figure}[t]
\centering
\includegraphics[width=\textwidth]{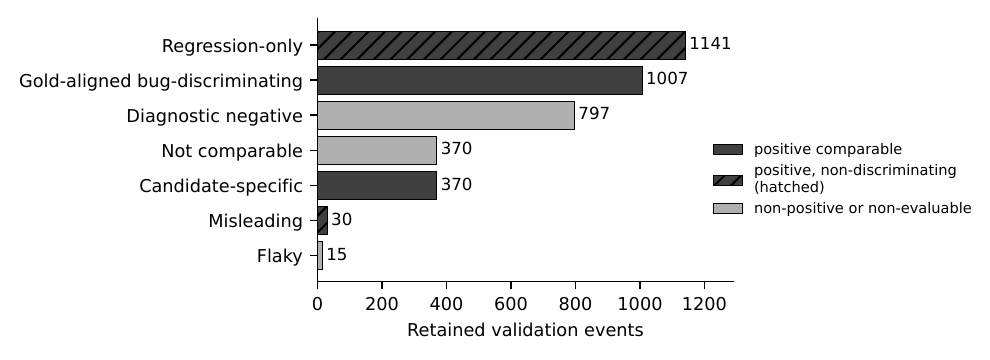}
\caption{Evidence roles of the \SQSFiveEventsTotal{} retained validation events
in the confirmatory experiment. Dark bars are positive comparable roles;
hatching marks positive roles that do not discriminate the reported bug}\label{fig:roles}
\end{figure}

Table~\ref{tab:taxonomy} and Figure~\ref{fig:roles} report the distribution of
evidence roles across \SQSFiveEventsTotal{} retained events from the
confirmatory experiment. Of these, \SQSFivePositiveComparable{} events are
positive comparable: the captured outcome was positive and the event holds one
of the four comparable roles. Within this set the roles divide as follows:
\SQSFiveRoleGoldAlignedBugDiscriminating{} gold-aligned bug-discriminating,
\SQSFiveRoleCandidateSpecific{} candidate-specific,
\SQSFiveRoleRegressionOnly{} regression-only, and \SQSFiveRoleMisleading{}
misleading. Regression-only and misleading events account for
\SQSFiveNondiscPositivePct\% of positive comparable events.

Among the \SQSFiveBfailSpass{} events where the check fails on B and passes on
the candidate, \SQSFiveCandSharePct\% are candidate-specific: they fail on the
developer gold fix G.

At the rollout level, \SQSFiveBaselineBdPct\% of baseline rollouts contain at
least one bug-discriminating event, while \SQSFiveBaselineEicPct\% exhibit
evidence-inadequate closure. Both patterns coexist: \SQSFiveCoexistPct\% of
baseline rollouts that submit a patch contain both discriminating and
non-discriminating positive evidence.

On the shared \SQCReplTasks-task subset of the second-model replication,
baseline \eic{} is \SQSSixbBaselineEicPct\% on gpt-5.6-terra versus
\SQSFiveSameTwoZeroBaselineEicPct\% on gpt-5.6-sol.

\begin{table}[t]
\caption{Evidence roles of the 3,730 retained validation events in the confirmatory experiment}\label{tab:taxonomy}
\begin{tabular}{lrr}
\toprule
Evidence role & Events & Share of events (\%) \\
\midrule
GOLD\_ALIGNED\_BUG\_DISCRIMINATING & 1,007 & 27.0 \\
CANDIDATE\_SPECIFIC & 370 & 9.9 \\
REGRESSION\_ONLY & 1,141 & 30.6 \\
MISLEADING & 30 & 0.8 \\
DIAGNOSTIC\_NEGATIVE & 797 & 21.4 \\
NOT\_COMPARABLE & 370 & 9.9 \\
FLAKY & 15 & 0.4 \\
\midrule
All retained events & 3,730 & 100.0 \\
Positive comparable events & 2,548 & 68.3 \\
\quad of which regression-only or misleading & 1,171 & 46.0\textsuperscript{a} \\
B-fail/S-pass events & 1,377 & 36.9 \\
\quad of which candidate-specific & 370 & 26.9\textsuperscript{b} \\
\botrule
\end{tabular}
\footnotetext{Shares in the two indented rows are relative to their parent row, marked \textsuperscript{a} of positive comparable events and \textsuperscript{b} of B-fail/S-pass events.}
\end{table}

\subsection{Evidence quality and repair success}\label{sec:rq2}

Over the \SQRqTwoN{} observed rollouts in
\SQRqTwoRepositories{} repositories (\SQRqTwoFolds{} repository-grouped
folds), the covariates-only model attains an out-of-fold Brier score of
\SQRqTwoBrierDesc{} against \SQRqTwoBrierFull{} for the
covariates-plus-evidence model, a reduction of
\SQRqTwoBrierReduction{} (repository-cluster bootstrap 95\% CI
\SQRqTwoBrierReductionCi); out-of-fold AUROC rises from \SQRqTwoAurocDesc{} to
\SQRqTwoAurocFull{}, and the augmented model recalibrates with intercept
\SQRqTwoCalibIntercept{} and slope \SQRqTwoCalibSlope{}. These
evidence terms carry associational predictive signal for official
resolution beyond the covariates. The analysis is associational and
predictive; it does not establish that evidence causes official resolution.

\subsection{Confirmatory experiment}\label{sec:confirmatory}

Of \SQDesignPlannedRollouts{} planned cells, \SQSFiveObservedRollouts{}
completed and \SQSFiveMissingRollouts{} were lost to infrastructure failures,
balanced across arms.

\eic{} (BCF minus Static Reminder) is \SQSFiveEicDiffPp{} percentage points
(95\% CI \SQSFiveEicCiPp, $p = \SQSFiveEicP$), based on \SQSFivePairedTasks{}
paired tasks. The difference in bug-discriminating evidence is
\SQSFiveBdDiffPp{} percentage points (95\% CI \SQSFiveBdCiPp,
$p = \SQSFiveBdP$). Both outcomes favour BCF in the prespecified fixed
sequence. Both point estimates fall below the prespecified
\SQDesignMcidAbsPp-percentage-point SESOI; the confidence intervals include the
SESOI but do not exclude effects well below it, so the data do not resolve
whether the true effects reach that threshold. Official resolution (BCF minus
Static Reminder) is
\SQSFiveResolvedDiffPp{} percentage points: BCF does not detectably change
repair success. Closure without positive machine-verifiable evidence shifts by
\SQSFiveNomvDiffPp{} points: BCF does not suppress validation activity.

Sign tests, leave-one-repository-out estimation, a repository-cluster wild
bootstrap \citep{CGM2008}, a complete-pairs restriction, and worst-case
missing-cell bounds all preserve the primary direction;
Appendix~\ref{app:robustness} reports them together with per-arm cell
accounting and trigger exposure (Table~\ref{tab:confirmatory}).

\subsection{Active-control decomposition}\label{sec:decomposition}

Table~\ref{tab:decomposition} decomposes the BCF effect into two components. The
generic reminder reduces \eic{} by \SQSFiveStaticEicAbsPp{} percentage points
relative to baseline and raises bug-discriminating evidence by
\SQSFiveStaticBdAbsPp{} points. BCF adds a further \SQSFiveEicDiffAbsPp{}
points on \eic{} and \SQSFiveBdDiffAbsPp{} points on bug-discriminating
evidence beyond the reminder. The total BCF-versus-baseline contrast is
\SQSFiveBcfbaseEicPp{} points on \eic{} and \SQSFiveBcfbaseBdPp{} on
bug-discriminating evidence.

\begin{table}[t]
\caption{Arm-level rates and three-arm decomposition of the confirmatory effects (task-level equal-source contrasts; Welch $p$ values outside the confirmatory family are descriptive)}\label{tab:decomposition}
\setlength{\tabcolsep}{3pt}
\begin{tabular}{lrrr}
\toprule
Rollout-level rate (\%) & Baseline & Static Reminder & BCF \\
\midrule
EIC & 23.8 & 21.4 & 13.6 \\
Bug-discriminating evidence & 72.9 & 75.8 & 83.2 \\
Official resolution & 65.9 & 66.0 & 66.8 \\
\midrule
Contrast & Difference (pp) & 95\% CI (pp) & $p$ \\
\midrule
Static $-$ Baseline, EIC & $-$3.22 & [$-$7.58, 1.13] & 0.15 \\
Static $-$ Baseline, bug-discriminating evidence & 4.12 & [$-$0.94, 9.17] & 0.11 \\
BCF $-$ Static, EIC & $-$7.79 & [$-$12.87, $-$2.72] & 0.0029 \\
BCF $-$ Static, bug-discriminating evidence & 7.36 & [1.73, 12.98] & 0.011 \\
BCF $-$ Baseline, EIC & $-$10.45 & [$-$15.78, $-$5.13] & 0.00018 \\
BCF $-$ Baseline, bug-discriminating evidence & 10.91 & [4.82, 17.00] & 0.00058 \\
\midrule
Secondary outcomes, BCF $-$ Static & & & \\
\midrule
Closure without positive machine-verifiable evidence & 0.44 & [$-$1.94, 2.82] & 0.72 \\
Official resolution & 0.48 & [$-$3.73, 4.69] & 0.82 \\
\botrule
\end{tabular}
\end{table}

\subsection{Runtime and token overhead}

Table~\ref{tab:resources} reports per-rollout resource means by arm. The
replay computation itself is inexpensive: BCF adds a median of
\SQSFiveBreplayMedianS{} seconds of B-replay time per rollout,
\SQSFiveBreplayWallPct\% of wall time at the median. The BCF-versus-Static
difference in per-rollout token consumption is \SQSFiveTokensDiff{} thousand
tokens, not statistically distinguishable from zero.

\begin{table}[t]
\caption{Runtime and token overhead of the interventions in the confirmatory experiment (per-rollout means over observed cells)}\label{tab:resources}
\begin{tabular}{lrrr}
\toprule
Per-rollout mean & Baseline & Static Reminder & BCF \\
\midrule
Model responses & 29.8 & 29.1 & 31.4 \\
Total tokens ($10^3$) & 956 & 946 & 1068 \\
Output tokens ($10^3$) & 10.7 & 10.5 & 11.5 \\
Wall time (s) & 390 & 381 & 440 \\
\botrule
\end{tabular}
\end{table}

\subsection{Exploratory replications}\label{sec:boundary}

\begin{figure}[t]
\centering
\includegraphics[width=\textwidth]{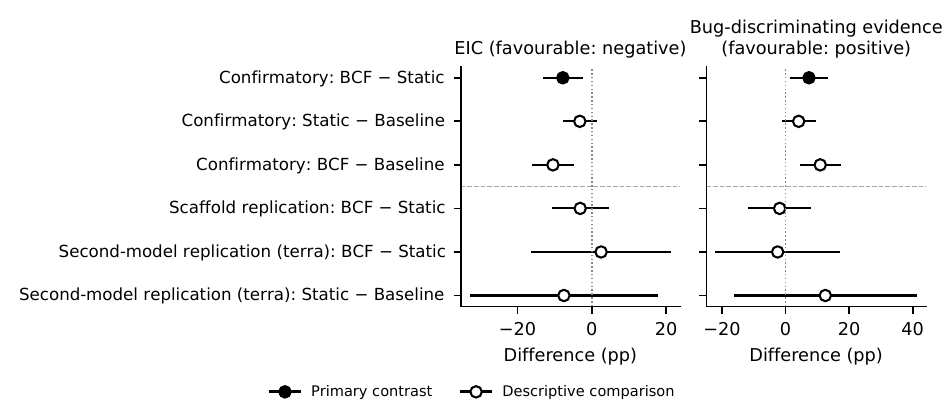}
\caption{Task-level equal-source differences with 95\% CIs across settings.
Filled markers show the primary contrast family; open markers show
descriptive comparisons. The dashed line
separates the confirmatory experiment from the exploratory replications}\label{fig:forest}
\end{figure}

Table~\ref{tab:replications} and Figure~\ref{fig:forest} report both
exploratory replications on the shared \SQCReplTasks-task subset;
Appendix~\ref{app:robustness} details cell loss and trigger exposure. The
scaffold replication reduces baseline evidence-inadequate closure to
\SQSSixBaselineEicPct\%, compared to \SQSFiveSameTwoZeroBaselineEicPct\% for
the same tasks under the unconstrained loop, leaving little room for either
component. On gpt-5.6-terra in the unconstrained loop, the generic reminder reduces
evidence-inadequate closure by \SQSSixbPlaceboEicAbsPp{} points relative to
baseline, matching the full BCF effect on sol, while the B-replay content adds
nothing detectable on top; feedback reached \SQSSixbTriggerPct\% of
intervention rollouts, so the null is not explained by too few intervention
triggers. Across the two studies, the reminder-only arm shows the more
consistent favourable pattern; an incremental BCF advantage is established
only with sol in the unconstrained loop.

\begin{table}[t]
\caption{Exploratory replications on the shared 20-task subset (BCF $-$ Static Reminder; all quantities are descriptive)}\label{tab:replications}
\begin{tabularx}{\textwidth}{>{\raggedright\arraybackslash}X *{3}{>{\centering\arraybackslash}p{2.45cm}}}
\toprule
 & Confirmatory (same 20 tasks) & Scaffold replication & Second-model replication (terra) \\
\midrule
Observed / planned cells & -- & 108/120 & 113/120 \\
EIC difference (pp) & $-$7.78 & $-$3.13 & 2.50 \\
EIC 95\% CI (pp) & -- & [$-$10.51, 4.26] & [$-$16.08, 21.08] \\
Bug-discriminating evidence difference (pp) & 7.78 & $-$1.88 & $-$2.50 \\
Bug-discriminating evidence 95\% CI (pp) & -- & [$-$11.56, 7.81] & [$-$21.74, 16.74] \\
Baseline EIC (\%) & -- & 8.3 & 40.5 \\
Baseline bug-discriminating evidence (\%) & -- & 83.3 & 51.4 \\
Interaction, EIC (pp; $p$) & -- & 5.00; 0.52 & 10.83; 0.24 \\
Interaction, bug-discriminating evidence (pp; $p$) & -- & $-$10.00; 0.33 & $-$10.56; 0.33 \\
Cells lost to failures (by arm) & -- & 12 (4/4/4) & 7 (3/2/2) \\
\botrule
\end{tabularx}
\footnotetext{Settings: the confirmatory experiment and the second-model replication run the unconstrained tool-use loop; the scaffold replication runs the plan-execute-verify scaffold. Models: gpt-5.6-sol for the confirmatory experiment and the scaffold replication; gpt-5.6-terra for the second-model replication. The confirmatory column restricts the confirmatory sample to the 19 of these 20 tasks with complete pairs. Interaction estimates compare each replication with the full confirmatory experiment and are descriptive.}
\end{table}

\section{Discussion}\label{sec:discussion}

\subsection{What the taxonomy reveals}

The central finding is a prevalence estimate: \SQSFiveNondiscPositivePct\% of
the positive validation evidence that agents produce mid-trajectory carries no
information about the reported defect. This holds in a population where
\SQSFiveBaselineBdPct\% of rollouts also contain at least one
discriminating event. Agents frequently produce both kinds of evidence in the
same run. The tool loop, however, exposes no signal that distinguishes the
two kinds: nothing indicates whether a passing check would also have passed
on B.

A finer-grained pattern sits within the discriminating events. Among events
where the check fails on B and passes on the candidate (the B-fail/S-pass
pattern), \SQSFiveCandSharePct\% also fail on the
developer gold fix G. Their discriminating power is tied to implementation
details of the agent's patch rather than to the defect itself: used as
acceptance tests downstream, they would reject correct alternative fixes.
This distinction is invisible without the G-replay that BSG-VA provides, and
it bears on the growing practice of using agent-generated tests as patch
validators \citep{SWT}.

On the shared subset of the second-model replication, baseline \eic{} is
\SQSSixbBaselineEicPct\% on gpt-5.6-terra versus
\SQSFiveSameTwoZeroBaselineEicPct\% on gpt-5.6-sol: the phenomenon is at
least as pronounced on gpt-5.6-terra.

\subsection{Awareness versus information}

We designed the three-arm experiment to distinguish two hypotheses about why
BCF works. The first is that agents simply need to be reminded to evaluate
their evidence. The second is that the specific B-replay content matters. The
generic reminder, matched in structure and timing to BCF but carrying no
replay information, serves as the attention control that separates these
contributions.

The results favour a mixed account. The reminder alone reduces \eic{} by
\SQSFiveStaticEicAbsPp{} percentage points relative to baseline: a nonspecific
prompting effect, captured by the attention-control arm. The B-replay content
adds a further \SQSFiveEicDiffAbsPp{} points beyond it. On sol in the
unconstrained loop, both components contribute. On terra, the prompting effect
alone accounts for the full improvement. Under the plan-execute-verify
scaffold, baseline \eic{} is already at \SQSSixBaselineEicPct\%, and neither
component has room to show an effect.

The practical takeaway is that the cheaper intervention has the broader reach.
A generic validation prompt, delivered at the right moment in the tool loop,
produces a favourable shift in both models we tested and requires no
replay infrastructure. The B-replay content adds incremental value with
gpt-5.6-sol in the unconstrained loop. Whether that increment justifies the
infrastructure cost depends on the deployment context. The measurement method
is relevant either way: it provides the means to evaluate evidence quality
regardless of whether a feedback loop is deployed.

\subsection{Threats to validity}\label{sec:limitations}

\paragraph{Construct validity.} Evidence roles are assigned against the
developer gold fix as reference standard, which introduces noise on tasks
where the gold fix is incomplete or where multiple valid fixes exist. The
\eic{} construct depends on the validation-event definition set before data
collection; alternative definitions of what counts as a validation event would
shift the prevalence estimates. The measurement begins at the agent's first
production edit, so diagnostic commands issued during initial bug exploration
fall outside the captured population. If agents produce discriminating checks
before editing code, the reported non-discriminating share would overstate the
trajectory-wide rate.

\paragraph{Internal validity.} Cells lost to infrastructure failures are balanced across
arms in all three experiments, and the confirmatory worst-case bound preserves the
sign of the effect; the second-model bound does not.

\paragraph{External validity.} Both models come from a single provider and
model family, and the task population draws on two Python-ecosystem
benchmarks. The exploratory replications cover a shared subset with few repetitions
per arm, yielding wide confidence intervals; their results are descriptive.

\section{Conclusion}\label{sec:conclusion}

We introduce BSG-VA, a method that measures the evidential value of the
validation activity repair agents perform mid-trajectory, by replaying each
command on the buggy, candidate, and gold-fix code states. Applied at scale,
the method reveals that \SQSFiveNondiscPositivePct\% of positive comparable
validation events carry no bug-discriminating information, and that
\SQSFiveBaselineEicPct\% of baseline rollouts close on the basis of such
evidence alone.

Bug-contrast feedback, which returns the B-replay outcome to the agent in real
time, reduces evidence-inadequate closure by \SQSFiveEicDiffAbsPp{} percentage
points, without measurable cost to repair
success. That estimate falls below the prespecified
\SQDesignMcidAbsPp-percentage-point SESOI, so the direction of the effect is
established while its practical magnitude is not.
The active-control decomposition attributes roughly a third of the
total improvement to directing the agent's attention to evidence quality, and
the exploratory replications show the same reminder-only pattern on the
second model; the richer B-replay content adds value in one
configuration, so a deployment without replay infrastructure still has a
measured, cheaper option in the generic reminder. What holds across every
configuration we tested is the measurement itself: BSG-VA and the
evidence-role taxonomy assign each passing check in any replayable trajectory an
evidence role with directly checkable content, making the adequacy of an
agent's validation evidence a property that can be audited before a patch is
trusted.

\backmatter

\section*{Statements and Declarations}

\textbf{Funding.} The authors received no funding for this work.

\textbf{Competing interests.} The authors have no competing interests to
declare.

\textbf{Ethics approval and consent to participate.} Not applicable. This
study involved no human participants and no animal subjects; all data are
machine-generated execution records.

\textbf{Consent for publication.} Not applicable.

\textbf{Data availability.} The event-level, rollout-level, and task-level
validation data supporting this study are openly available on Zenodo at
\url{https://doi.org/10.5281/zenodo.21642576} \citep{xu2026validation}. The release
carries the three JSONL files behind every reported statistic. Replay
environments are pinned by public container image references. The protocol
documents, field definitions and comparability rules, and the raw trajectory
logs are available from the corresponding author on reasonable request.

\textbf{Code availability.} The BSG-VA capture, replay, and analysis code
used to produce every reported statistic is not publicly released; it is
available from the corresponding author on reasonable request.

\textbf{Author contributions.} Conceptualization: Xiaonan Xu. Methodology:
Xiaonan Xu. Formal analysis: Xiaonan Xu. Software: Wenjing Wu. Data curation:
Wenjing Wu. Writing, original draft: Xiaonan Xu. Writing, review and editing:
Xiaonan Xu and Wenjing Wu. Both authors read and approved the final
manuscript.

\textbf{Use of AI tools.} Section~\ref{sec:ai-use} details the role of an LLM
assistant (Claude, Fable~5 model, Anthropic) in environment configuration,
experiment implementation, analysis code, and language polishing. The authors
take responsibility for all content.

\begin{appendices}

\section{Robustness analyses and exposure details}\label{app:robustness}

\subsection{Confirmatory robustness}

Table~\ref{tab:confirmatory} collects the robustness checks for the primary
contrast. The exact sign test on task-level differences is
\SQSFiveSignEicBetter{}:\SQSFiveSignEicWorse{} for \eic{}
($p = \SQSFiveSignEicP$) and \SQSFiveSignBdBetter{}:\SQSFiveSignBdWorse{} for
bug-discriminating evidence ($p = \SQSFiveSignBdP$). Leave-one-repository-out
estimation is favourable in \SQSFiveLoroEicFav/\SQSFiveLoroRepos{}
repositories for \eic{} and \SQSFiveLoroBdFav/\SQSFiveLoroRepos{} for
bug-discriminating evidence. The repository-cluster wild bootstrap
(\SQSFiveWildDraws{} draws) yields $p = \SQSFiveEicWildP$ for \eic{} and
$p = \SQSFiveBdWildP$ for bug-discriminating evidence. Restricting to the
\SQSFiveCompletePairsTasks{} tasks with complete pairs in both intervention
arms gives \SQSFiveEicCompletePairsDiffPp{} points on \eic. The
\SQSFiveMissingRollouts{} cells lost to infrastructure failures split
\SQSFiveMissingBaseline/\SQSFiveMissingStatic/\SQSFiveMissingBcf{} across
baseline, Static Reminder, and BCF; under worst-case imputation of
all of them, the \eic{} difference attenuates to \SQSFiveEicAdverseBoundPp{}
points and the bug-discriminating-evidence difference to
\SQSFiveBdAdverseBoundPp{}, preserving direction in both cases.

\begin{table}[t]
\caption{Confirmatory contrast (BCF versus Static Reminder, 109 paired tasks, equal-source weighting) and robustness analyses}\label{tab:confirmatory}
\begin{tabular}{lcc}
\toprule
 & EIC (pp) & Bug-discriminating evidence (pp) \\
\midrule
Task-level difference & $-$7.79 & 7.36 \\
Welch 95\% CI & [$-$12.87, $-$2.72] & [1.73, 12.98] \\
Welch two-sided $p$ & 0.0029 & 0.011 \\
Repository wild bootstrap $p$ & 0.00020 & 0.0018 \\
Wild bootstrap 95\% CI & [$-$12.16, $-$3.43] & [2.55, 12.13] \\
Exact sign test (favourable:unfavourable) & 19:5 & 21:7 \\
Exact sign test $p$ & 0.0066 & 0.013 \\
Leave-one-repository-out favourable & 63/63 & 63/63 \\
Complete-pairs difference (100 tasks) & $-$8.00 & 7.45 \\
Worst-case missingness bound & $-$5.00 & 4.55 \\
\botrule
\end{tabular}
\footnotetext{Fixed-sequence family: EIC tested first at two-sided $\alpha = 0.05$; bug-discriminating evidence tested only after EIC rejection in its favourable direction. Additional analyses are reported as robustness checks.}
\end{table}

\subsection{Trigger exposure}

Feedback reached \SQSFiveTriggerStaticPct\% of Static Reminder rollouts and
\SQSFiveTriggerBcfPct\% of BCF rollouts in the confirmatory experiment. In the
second-model replication the trigger rate is \SQSSixbTriggerPct\% in both
intervention arms.

\subsection{Replication cell loss}

The scaffold replication lost \SQSSixMissingRollouts{} of
\SQCReplPlannedCells{} cells (\SQSSixMissingArmSplit{} by arm);
\SQSSixPygmtCells{} of these are one task whose rollouts the scaffold's
submission gate aborted, affecting all arms equally (two cells per arm). The
second-model replication lost \SQSSixbMissingRollouts{} cells
(\SQSSixbMissingArmSplit{} by arm) to infrastructure failures. The
second-model worst-case missingness bound can reverse the sign of the
estimate.

\end{appendices}



\end{document}